\documentclass[reprint,amsmath,amssymb]{revtex4-1}
\usepackage[utf8x]{inputenc}
\usepackage{graphicx}
\usepackage{dcolumn}
\usepackage{bm}
\usepackage{color}

\usepackage{soul} 
\usepackage[normalem]{ulem} 

\begin{document}

\preprint{APS/123-QED}

\title{Superfluid weight and polarization amplitude in the
  one-dimensional bosonic Hubbard model}

\author{B. Het\'enyi$^{1,2}$, L.~M. Martelo$^{3,4}$, and
  B. Tanatar$^1$} \affiliation{$^1$ Department of Physics, Bilkent
  University, TR-06800 Bilkent, Ankara, Turkey \\ $^2$MTA-BME Quantum
  Dynamics and Correlations Research Group, Department of Physics,
  Budapest University of Technology and Economics, H-1111 Budapest,
  Hungary \\ $^3$Departamento de Engenharia F\'isica, Faculdade de
  Engenharia da Universidade do Porto, Rua Dr. Roberto Frias, 4200-465
  Porto, Portugal \\ $^4$Centro de F\'isica do Porto, Faculdade de
  Ciências da Universidade do Porto, Rua do Campo Alegre, 687,
  4169-007 Porto, Portugal}

\date{\today}

\begin{abstract}
\textcolor{black}{We calculate the superfluid weight and the
  polarization amplitude for the one-dimensional bosonic Hubbard model
  with focus on the strong-coupling regime via variational, exact
  diagonalization, and strong coupling calculations.  Our variational
  approach is based on the Baeriswyl wave function, implemented via
  Monte Carlo sampling.  We derive the superfluid weight appropriate
  in a variational setting.  We emphasize the importance of
  implementing the Peierls phase in position space and to allow for
  many-body interference effects, rather than implementing the Peierls
  phase as single particle momentum shifts.  At integer filling, the
  Baeriswyl wave function gives zero superfluid response at any
  coupling.  At half-filling our variational superfluid weight is in
  reasonable agreement with exact diagonalziation results.  We also
  calculate the polarization amplitude, the variance of the total
  position, and the associated size scaling exponent, which
  corroborate that this variational approach produces an insulating
  state at integer filling.  Our Baeriswyl based variational method is
  applicable to significantly larger system sizes than exact
  diagonalization or quantum Monte Carlo.}
\end{abstract}


\maketitle

\section{Introduction}

The bosonic Hubbard model (BHM) was introduced by Gersch and
Knollman~\cite{Gersch63} to study the condensation of repulsive
interacting bosons.  The phase diagram of the superfluid-insulator
transition was described by Fisher {\it et al.}~\cite{Fisher89}.
Since then the phase diagram has been calculated and refined by a
variety of means, including mean-field
theory~\cite{Fisher89,Rokhsar91}, perturbative
expansion~\cite{Freericks96}, quantum Monte
Carlo~\cite{Batrouni90,Krauth91,Scalettar05,Rousseau06,Kashurnikov96b,Rombouts06}
(QMC), density matrix renormalization
group~\cite{Kuhner00,Ejima11,Zakrzewski08,Carrasquilla13,Kuhner98},
and exact diagonalization~\cite{Kashurnikov96a}.  For a review see
Ref. ~\cite{Pollet12}.  Due to the experimental realization of the
model~\cite{Jaksch98,Greiner02} as ultracold atoms in an optical
lattice, the model has gained renewed interest.

The BHM is often applied to model Bose solids, e.g. solid
$^4$He~\cite{Anderson12}.  One question of interest in these systems
is under what conditions a superfluid type response is
exhibited~\cite{Reatto88,Vitiello88}.  Some experimental~\cite{Kim06}
and some theoretical~\cite{Galli05} results suggest that solid helium
becomes supersolid at low temperatures.  The experimental conclusions,
some of which are based on torsional oscillator measurements, have
since been challenged by the suggestion that other effects may be
responsible for the observed drop in rotational inertia, such as
quantum plasticity~\cite{Beamish10}, moreover the role played by
defects still awaits clarification (see Ref. \cite{Hallock15} for an
overview).  For the BHM Anderson~\cite{Anderson12} has suggested that
the ground state at integer filling is a supersolid.

In this work we apply a variational Monte Carlo~\cite{Hetenyi16} for
strongly correlated bosonic models based on the Baeriswyl
wavefunction~\cite{Baeriswyl86,Baeriswyl00,Dzierzawa97,Martelo97,Hetenyi10,Dora15,Yahyavi18}
(BWF) and exact diagonalization (ED) to study the superfluid response
and the polarization
amplitude~\cite{Resta98,Resta99,Aligia99,Souza00,Nakamura02,Yahyavi17,Kobayashi18,Hetenyi19,Nakamura19,Furuya19}
of the one-dimensional BHM.  \textcolor{black}{Central to our study is
  the derivation of the expression for the superfluid weight valid in
  variational calculations, and emphasis of the role of interference
  between Peierls phases of particle paths in calculating the
  superfluid weight.  According to our derivation, the Peierls phases
  should be implemented in position space, and phases along the paths
  of different particles should be allowed to interfere between
  exchanging particles.  If the relevant propagators are Fourier
  transformed, and the Peierls phases are implemented as
  single-particle momentum shifts, interference effects are missed.
  The Fourier transform of the polarization amplitude shows a peak at
  small hopping strength for both the ED and the BWF-MC, which both
  decrease as the hopping strength is increased.  This similarity is
  merely quantitative, the insulator-superfluid transition is only
  picked up in ED, where the scaling exponent of the variance of the
  polarization indicates gap closure.  We also analyze Anderson's
  treatment~\cite{Anderson12} of the BHM in light of our findings.}

Our paper is organized as follows.  In the next section we present the
model and the two methods used in this work, ED and the BWF-MC method.
Subsequently, we discuss the superfluid weight. In section
\ref{sec:polarization} the polarization amplitude and cumulants are
described.  In section \ref{sec:results} our results are presented.
In the penultimate section we comment on the calculation of the
superfluid weight by Anderson, subsequently we conclude our work.  A
strong coupling treatment is presented in the Appendix.

\section{Model and methods}

\label{sec:modelmethods}

We study the BHM with nearest neighbor hopping in one dimension at
fixed particle number.  The Hamiltonian is
\begin{equation}
\label{eqn:hml}
H = -J\sum_{i=1}^L \left(\hat{c}^\dagger_{i+1} \hat{c}_i + \hat{c}^\dagger_{i} \hat{c}_{i+1}\right) + U \sum_{i=1}^L \hat{n}_i (\hat{n}_i-1),
\end{equation}
where $L$ denotes the number of sites, $J$ and $U>0$ are the hopping
and repulsive on-site interaction parameters, respectively,
$\hat{c}^\dagger_i$($\hat{c}_i$) denotes a bosonic
creation(annihilation) operator at site $i$, and $\hat{n}_i$ denotes
the density operator at site $i$.

\textcolor{black}{While the method we developed is described in detail for the BWF
elsewhere~\cite{Hetenyi16}, here we describe aspects that are relevant
to this study.  The BWF has the form
\begin{equation}
\label{eqn:PsiBWF}
|\Psi_B\rangle = \exp(-\alpha\hat{T})|\infty\rangle,
\end{equation}
where $\alpha$ denotes the variational parameter, and $|\infty\rangle$
is the wavefunction at $U=\infty$.  $\hat{T}$ denotes the hopping
operator (first term in Eq. (\ref{eqn:hml})).  The crucial step in the
construction of our method is that the kinetic energy propagator can
be represented in real space as,
\begin{equation}
\label{eqn:prpt}
\langle x  | \exp(-\alpha\hat{T})  | x' \rangle = 
\frac{1}{L}\sum_k e^{-\alpha \epsilon_k + i k (x-x')},
\end{equation}
where $\epsilon_k = -2J\cos(k)$.  The full probability sampled~\cite{Hetenyi16}
consists of a {\it product} of single particle propagators.  Bosonic
exchanges have to be implemented by exchanging the positions of pairs
of particles and accepting or rejecting such exchange moves.}

\begin{figure}[ht]
 \centering
 \includegraphics[width=8cm,keepaspectratio=true]{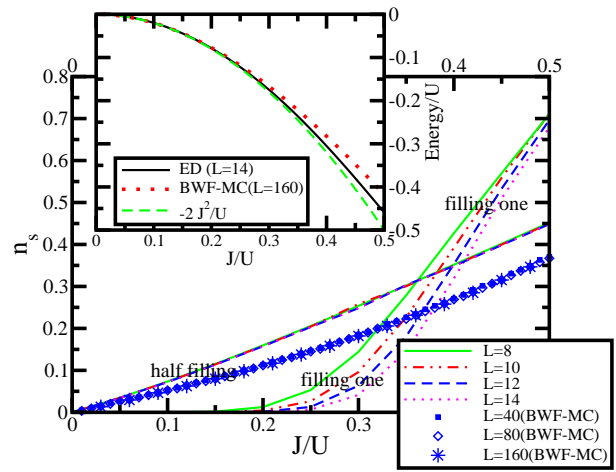}
 \caption{Superfluid weight based on exact diagonalization
   calculations and BWF-MC as a function of $J/U$ for different system
   sizes for fillings of one-half and one.  The inset compares the
   energy per particle for integer filling for the Baeriswyl wave
   function based variational Monte Carlo method, exact
   diagonalization, and the strong coupling expansion based on the
   Baeriswyl wave function ($-2J^2/U$) for filling one.}
 \label{fig:ns_ED}
\end{figure}

\textcolor{black}{For smaller systems, we diagonalize Eq. (\ref{eqn:hml}) by the
well-known iterative L\'anczos scheme.  An important aspect of the
method we use is the indexing of the states, which is based on the
Lehmer combinatorial code~\cite{Lehmer60,Laisant88}, a way to order
permutations.  This procedure is also called Ponomarev
ordering~\cite{Ponomarev09,Ponomarev10,Ponomarev11} and has been
implemented in the BHM by Raventos {\it et al.}~\cite{Raventos17}.}

\begin{figure}[ht]
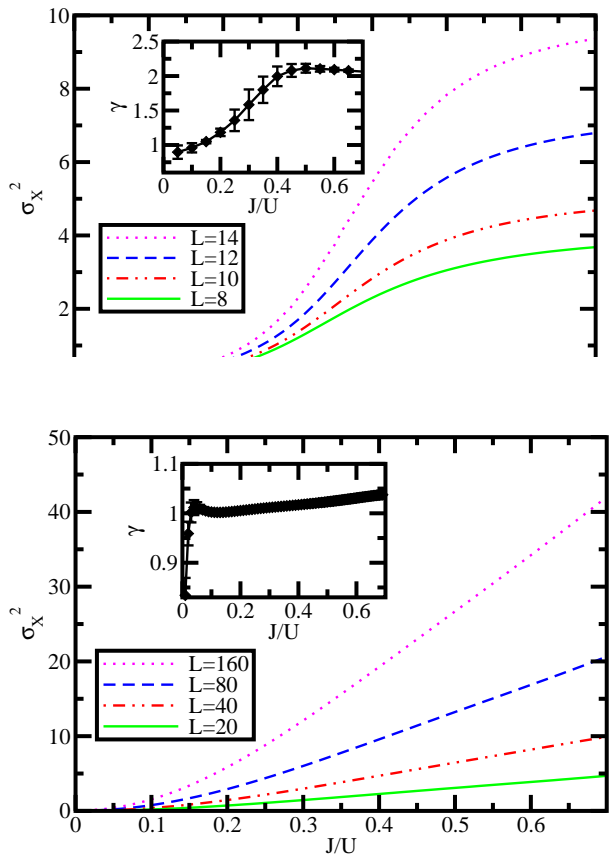

 \centering
 \includegraphics[width=8cm,keepaspectratio=true]{./sigX2_ED.eps}
 \includegraphics[width=8cm,keepaspectratio=true]{./sigX2_BWFMC.eps}
 \caption{Second cumulant of the polarization as a function of $J/U$
   for filling one for different system sizes based on exact
   diagonalization (upper panel) and variational Monte Carlo (lower
   panel) calculation.  The inset shows the respective size scaling
   exponents.  }
 \label{fig:sigX2}
\end{figure}

\section{Superfluid weight}

\label{sec:SFW}
 
It is not immediately obvious that
there is an issue with this quantity when it is considered
specifically in a variational context.  Usually the second derivative
of the variational ground state energy with respect to the flux at
zero Peierls flux is calculated~\cite{Millis91} (see
Eq. (\ref{eqn:hml_Phi}) for how the Peierls phase enters).  One way to
elucidate the issue~\cite{Hetenyi14} is to consider that a variational
estimate of the ground state energy is a weighted average of exact
energy eigenvalues,
\begin{equation}
  \label{eqn:PE}
  E_{var} = \sum_i P_i(\Phi) E_i(\Phi),
\end{equation}
where $P_i = |\langle \Psi(\Phi) | \psi_i(\Phi) \rangle|^2$
($|\Psi(\Phi)\rangle$ denotes the variational wave function,
$|\psi_i(\Phi) \rangle$ denotes the $i$th eigenstate of the
Hamiltonian), and $E_i(\Phi)$ denotes the $i$th energy eigenvalue.  We
argue here that the correct superfluid weight, when calculated in a
variational calculation, is given by the expression
\begin{equation}
  \label{eqn:n_S_v}
  n_S = \frac{1}{L}\left.\sum_i P_i(\Phi) \frac{\partial^2 E_i (\Phi)}{\partial \Phi^2}\right|_{\Phi=0},
\end{equation}
in other words, the derivative of $P_i(\Phi)$ with respect to the flux
need not be taken, in spite of its $\Phi$-dependence.  However, in
actual variational calculations, often neither $E_i(\Phi)$, nor
$P_i(\Phi)$ are available, so we give an alternative, but equivalent
expression for $n_S$, applicable in numerical settings.

Let us briefly recall the arguments of Pollock and
Ceperley~\cite{Pollock87}.  In their work a continuous system of
interacting atoms is considered, at finite temperature.  Below we
modify their steps to account for our lattice model.  Pollock and
Ceperley~\cite{Pollock87} considered a thought experiment, aimed to
mimic the rotating bucket experiments on superfluids.  In this setup,
the sample is rotated and the rotational inertia is measured.  Below
the critical temperature, where the superfluid fraction ceases to
rotate with the container, the rotational inertia takes a
non-classical value~\cite{Fisher73,Rousseau14}, different from the
rotational inertia calculated from the amount of fluid present in the
container.  In the context of supersolidity Leggett
suggested~\cite{Leggett70} that torsional oscillator experiments can
access the rotational inertia, and therefore the superfluid weight.

In Ref. \cite{Pollock87} the sample is enclosed between a two circular
walls, one of radius $R$, the other of radius $R+d$.  When $R\gg d$, the
system becomes equivalent to the sample between two parallel planes.
In the experiment the walls are moved by an outside agent with
velocity $v$.  It is expected that the normal component of the fluid
will move with the walls, due to friction, while the superfluid
component will remain stationary in the laboratory frame.

The density matrix of the system is
\begin{equation}
  \hat{\rho}_v = \exp(-\beta \hat{H}_v), 
\end{equation}
where $\beta$ denotes the inverse temperature, and
\begin{equation}
  \hat{H}_v = \sum_{j=1}^N \frac{(\hat{p}_j - m v)^2}{2m} + \hat{V},
\end{equation}
where $\hat{p}_j$ denotes the momentum of an individual particle, $m$
denotes the mass of a particle, and $\hat{V}$ denotes an interaction.
The total momentum of the system (the momentum of the normal
component) is given by
\begin{equation}
  \label{eqn:rho_n}
  \frac{\rho_n}{\rho} N m v = \frac{\mbox{Tr}\{\hat{\rho}_v \hat{P}\}}{\mbox{Tr}\{\hat{\rho}_v \}},
\end{equation}
where $\rho_n$ denotes the density of the normal phase, $\rho$ denotes
the total density, and 
\begin{equation}
  \label{eqn:P}
  \hat{P} = \left.\frac{\partial{\hat{H}_v}}{\partial v}\right|_{v=0} = \sum_{j=1}^N \hat{p}_j.
\end{equation}
We can write the free energy of the system as
\begin{equation}
  \exp(-\beta F_v) = \mbox{Tr} \{ \exp(-\beta \hat{H}_v)\}.
\end{equation}
Taking the first derivative of $F_v$ with respect to $v$ results in
\begin{equation}
  \label{eqn:delFv}
  \frac{\partial F_v}{\partial v} = N m v \left( 1 - \frac{\rho_n}{\rho}\right),
\end{equation}
resulting in the superfluid weight of
\begin{equation}
  \label{eqn:rho_s}
\frac{\rho_s}{\rho} = \frac{1}{N m} \left.\frac{\partial^2 F_v}{\partial v^2}\right|_{v=0}.
\end{equation}

In a lattice model the analog of the velocity $v$ is the Peierls
phase, which we introduce into the Hamiltonian as
\begin{equation}
\label{eqn:hml_Phi}
H(\Phi) = -J\sum_{i=1}^L \left(e^{i \Phi}\hat{c}^\dagger_{i+1}
\hat{c}_i + e^{-i \Phi}\hat{c}^\dagger_{i} \hat{c}_{i+1}\right) + U
\sum_{i=1}^L \hat{n}_i (\hat{n}_i-1).
\end{equation}
In exact diagonalization calculations, the analog of
Eq. (\ref{eqn:rho_s}), the superfluid weight, is obtained by
calculating the response of the system to the Peierls phase (or
boundary twist) as,
\begin{equation}
\label{eqn:ns}
n_S = \frac{1}{L}\left[ \frac{\partial^2 E_g(\Phi)}{\partial \Phi^2}
  \right]_{\Phi=0}.
\end{equation}
Here $E_g(\Phi)$ denotes the ground state of the system, since our
lattice system is studied at zero temperature.

In the BWF-MC method we use, where the exact eigenenergies are not
available, it is not possible to calculate the superfluid weight
directly (via Eq. (\ref{eqn:ns})).  Instead we derive the superfluid
weight in a variational context based on the reasoning of Pollock and
Ceperley~\cite{Pollock87}.  Our starting point is the normalization of
the Baeriswyl wavefunction,
\begin{equation}
  \label{eqn:NPhi}
  \tilde{Q}_\Phi = \langle \infty | \exp \left( - 2 \alpha
  \hat{T}(\Phi) \right) | \infty \rangle = \exp(-2\alpha \tilde{F}_\Phi),
\end{equation}
which is analogous to the partition function in statistical mechanics.
We also defined $\tilde{F}_\Phi$, the ``free energy'' in the
variational context.  The analog of Eq. (\ref{eqn:delFv}) is
\begin{equation}
\frac{\partial \tilde{F}_\Phi}{\partial \Phi} = \cos \Phi J_\Phi -
\sin \Phi T_\Phi,
\end{equation}
where
\begin{equation}
  \label{eqn:J_Phi}
  J_\Phi = \langle \infty | \exp \left( - \alpha
  \hat{T}(\Phi) \right)\hat{J}\exp \left( - \alpha
  \hat{T}(\Phi) \right)
  | \infty \rangle,
\end{equation}
with the current operator $\hat{J}$ being
\begin{equation}
  \label{eqn:J}
\hat{J} = \left.\frac{\partial H(\Phi)}{\partial \Phi} \right|_{\Phi=0} = -2J
  \sum_k \sin(k) n_k,
\end{equation}
and
\begin{equation}
  \label{eqn:T_Phi}
  T_\Phi = \langle \infty | \exp \left( - \alpha
  \hat{T}(\Phi) \right)\hat{T}\exp \left( - \alpha
  \hat{T}(\Phi) \right)
  | \infty \rangle,
\end{equation}
with $\hat{T}$ indicating the hopping energy operator at zero flux.
Note that $\hat{J}$ is the lattice analog of the total momentum
operator (compare Eq.(\ref{eqn:J}) and Eq. (\ref{eqn:P})).

From Eq. (\ref{eqn:delFv}), we derive our proposed expression for the
superfluid weight for variational calculations,
\begin{equation}
  \label{eqn:n_S}
n_S = \frac{1}{L}  \left.\frac{\partial^2 \tilde{F}_\Phi}{\partial \Phi^2}\right|_{\Phi=0} = \frac{1}{L}\left(
  \frac{\partial J_0}{\partial \Phi} - T_0\right).
\end{equation}
In the case of an exact eigenstate, this expression corresponds to
Eq. (\ref{eqn:ns}), but in the case of a weighted average of the type
in Eq. (\ref{eqn:PE}) it corresponds to Eq. (\ref{eqn:n_S_v}).  To see
this, we first write $T_0$ as
\begin{equation}
  T_0 = \sum_i P_i T_i
\end{equation}
where $T_i$ denotes the expectation value of the kinetic energy in
eigenstate $i$, and write $J_\Phi$ as
\begin{equation}
  J_\Phi = \sum_i P_i(\Phi) J_i(\Phi), 
\end{equation}
where $J_i(\Phi)$ denotes the expectation value of the current
operator ($\hat{J}$) in the eigenstate $|\psi_i(\Phi)\rangle$, and
take the derivative with respect to $\Phi$, resulting in
\begin{equation}
  \frac{\partial J_\Phi}{\partial \Phi} = \sum_i \left[
  \frac{\partial P_i(\Phi)}{\partial \Phi}  J_i(\Phi)
  +
  P_i(\Phi) \frac{\partial J_i(\Phi)}{\partial \Phi}
  \right].
\end{equation}
Setting $\Phi=0$, we can use the fact that $J_i(0)=0$, leaving us with
\begin{equation}
n_S = \frac{1}{L} \sum_i P_i \left( \frac{\partial J_i(0)}{\partial \Phi}
  - T_i\right),
\end{equation}
which is exactly  Eq. (\ref{eqn:n_S_v}).

The second derivative in Eq. (\ref{eqn:n_S}) involves two limits,
$\Phi \rightarrow 0$ and $L \rightarrow \infty$.  In one dimension the
order of limits is inconsequential, the superfluid weight is obtained
in both cases~\cite{Scalapino92,Scalapino93}.  Below we evaluate the
second derivative in Eq. (\ref{eqn:n_S}) by way of finite difference
on a grid $\Phi = m 2\pi/L$, with $m$ integer.

Some care needs to be exercised in applying the derivative with
respect to $\Phi$.  It appears highly tempting to implement the
Peierls phase as a shift in $k$, as $\epsilon_k \rightarrow
\epsilon_{k+\Phi}$ in each single particle propagator
(Eq. (\ref{eqn:prpt})).  This approach leads to a finite superfluid
weight even at integer filling.  However, in this case the
interference between the particles is neglected.

To elaborate, let us write $\tilde{Q}_\Phi$ in the coordinate
representation as
\begin{equation}
  \tilde{Q}_\Phi = \sum_{\bf x_L,x_R}\langle \infty | {\bf x_L} \rangle
  \langle {\bf x_L} | \exp \left( -2 \alpha \hat{T}(\Phi) \right) |{\bf x_R} \rangle \langle {\bf x_R} | \infty \rangle,
\end{equation}
where ${\bf x_{L/R}} = x_{L/R,1},...,x_{L/R,N}$, indicating
particle positions in the ``left'' or ``right'' coordinate bases.  For
the moment, let us consider a one-particle propagator,
\begin{equation}
  \langle x | \exp \left( -2 \alpha \hat{T}(\Phi) \right) |x' \rangle =
  \langle x | (1 - 2 \alpha \hat{T}(\Phi) + 2 \alpha^2 \hat{T}(\Phi)\hat{T}(\Phi)) |x' \rangle
\end{equation}
A term in which the difference between $x'$ and $x$ is a fixed number
will involve all different paths which go from $x$ to $x'$.  The
different paths may have a different number of hoppings, left and
right, but the difference between $x'$ and $x$ is the same.  Each
right hopping contributes a phase of $\Phi$, while each left hopping
contributes a phase of $-\Phi$.  Thus, the net change in phase will be
determined solely by $x'-x$, we can write,
\begin{eqnarray}
  \langle x | \exp \left( -2 \alpha \hat{T}(\Phi) \right) |x' \rangle = &
  \exp\left[ i \Phi (x'-x) \right] \times \\
  \nonumber & \langle x | \exp \left( -2 \alpha \hat{T}(0) \right) |x' \rangle.
\end{eqnarray}
We used the fact that for a periodic system $\langle x | x' \rangle =
\delta_{x,x'+L}$.  Armed with this, we can rewrite $\tilde{Q}(\Phi)$
as
\begin{eqnarray}
  \tilde{Q}_\Phi = \sum_{\bf x_L,x_R}
  \exp\left[ i \Phi \sum_{i=1}^N (x_{R,i} - x_{L,i}) \right] \times \\ \nonumber
\hspace{.5cm}  \langle \infty | {\bf x_L} \rangle
  \langle {\bf x_L} | \exp \left( -2 \alpha \hat{T}(0) \right) |{\bf x_R} \rangle \langle {\bf x_R} | \infty \rangle.
\end{eqnarray}
We note that the operator appearing in the exponential $\sum_{i=1}^N
(x_{R,i} - x_{L,i})$ is a sum of differences between single particle
positions.  The contribution from different particles are now added,
and they can cancel.  The position operator is undefined in a periodic
system, but its exponential, provided that $\Phi = m 2 \pi/L$ is
well-defined.  Such an operator is used in the many-body analog of the
modern theory of polarization~\cite{Resta98,Resta99}, also to express
a second derivative in the momentum shift~\cite{Hetenyi19}.  Following
the same steps, we can write
\begin{equation}
  n_S = - \lim_{L \rightarrow \infty} \left[ \frac{L^2}{4 \alpha \pi^2} \right] \mbox{Re} \ln \tilde{Q}_{2 \pi / L}
\end{equation}
If the system has integer filling, it always holds that
\begin{equation}
  \sum_{i=1}^N (x_{R,i} - x_{L,i}) = 0,
\end{equation}
leading to a superfluid weight of zero.  According to this derivation
a finite superfluid weight can only arise for non-integer fillings.
This coincides with previous results on the Drude weight of the
Baeriswyl wave function in fermionic systems~\cite{Dzierzawa97}.

The derivation we provided above is specific to the BWF, in which the
hopping energy appears explicitly in the projector.  In variational
wave functions where that is not the case, one has to apply a
Baeriswyl projector, threaded by a Peierls flux, and take the limit of
$\alpha \rightarrow 0$ in the final expression for $n_S$.

\section{Polarization amplitude}

\label{sec:polarization}
 
For a periodic lattice system the total position operator is
ill-defined.  In a many-body system the standard approach
~\cite{Resta98} is to calculate the expectation value of the twist
operator~\cite{Lieb61}, also known as the polarization
amplitude~\cite{Kobayashi18},
\begin{equation}
  Z_q = \langle \Psi |\exp \left( i \frac{2 \pi q }{L} \hat{X}\right)|
  \Psi \rangle.
\end{equation}
where $\hat{X} = \sum_{i=1}^N x_i\hat{n}_i$. This quantity can be
interpreted as a characteristic function, and derivatives with respect
to $q$ give the average of $\hat{X}$, the variance of $\hat{X}$ or
higher order cumulants.  The second cumulant, $\sigma^2_N$, which
measures the spread of the center of mass, can be used to determine
whether a system is a conductor or an
insulator~\cite{Kohn64,Resta98,Resta99,Souza00,Yahyavi17}. In a finite
system the definition of the spread is
\begin{equation}
\label{eqn:totX}
\sigma^2_N = -\frac{L^2}{4\pi^2} \left. \frac{ \Delta^2 \ln
  Z_q}{\Delta q^2}\right|_{q=0},
\end{equation}
where $\frac{\Delta G}{\Delta q}$ indicates a discrete derivative of
$G$ with respect to $q$.  While the commonly used expression for the
variance of the total position of Resta and Sorella
is~\cite{Resta98,Resta99}, $\sigma^2 = -\frac{L^2}{2\pi^2} \mbox{Re}
\ln Z_1,$ (consistent with Eq. (\ref{eqn:totX})), for small system
sizes scaling exponents are difficult to obtain from
Eq. (\ref{eqn:totX}).  A simple remedy~\cite{Hetenyi19} is to take the
derivative of the ln function analytically, and apply discrete
derivatives to the remaining cases,
\begin{equation}
\label{eqn:totX2}
\sigma^2_N = -\frac{L^2}{4\pi^2}
\left[\left(
  \left. \frac{ \Delta^2 Z_q}{\Delta q^2}\right|_{q=0}\right)
  -
\left(\left. \frac{ \Delta Z_q}{\Delta q}\right|_{q=0}\right)^2
\right].
\end{equation}
In our calculations below we will use Eq. (\ref{eqn:totX2}) to
calculate the variance and the size scaling exponent $\gamma$ from an
assumed scaling form of
\begin{equation}
  \sigma^2_N(L)=aL^\gamma.
\end{equation}
We also calculate the discrete Fourier transform of $Z_q$, which we
define as
\begin{equation}
  \tilde{Z}_x = \sum_{q=1}^L \exp\left(i \frac{2 \pi q x}{L}\right) Z_q,
\end{equation}
where $x = 1,...,L$.

\begin{figure}[ht]
 \centering
 \includegraphics[width=8cm,keepaspectratio=true]{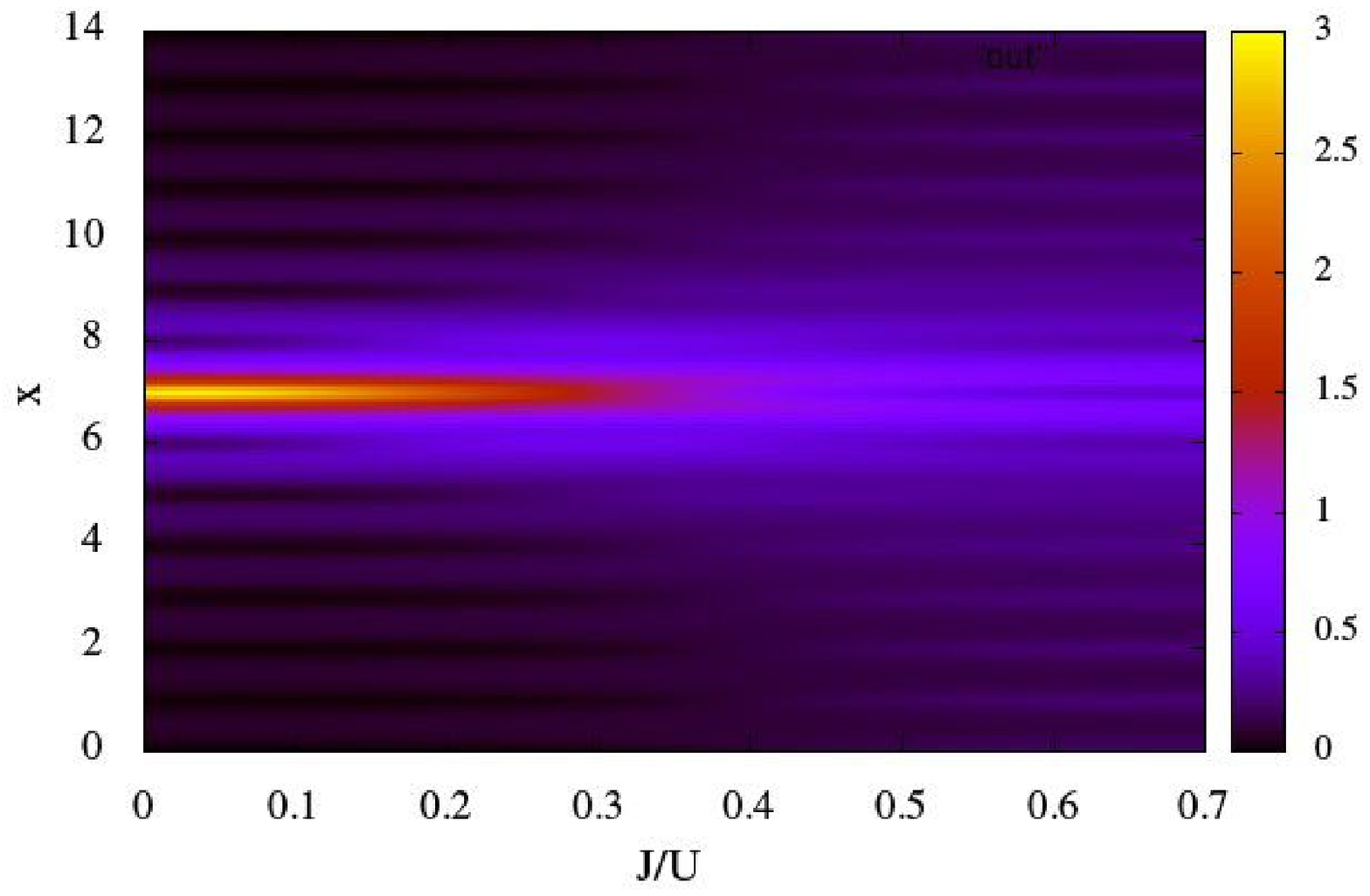}
 \includegraphics[width=8cm,keepaspectratio=true]{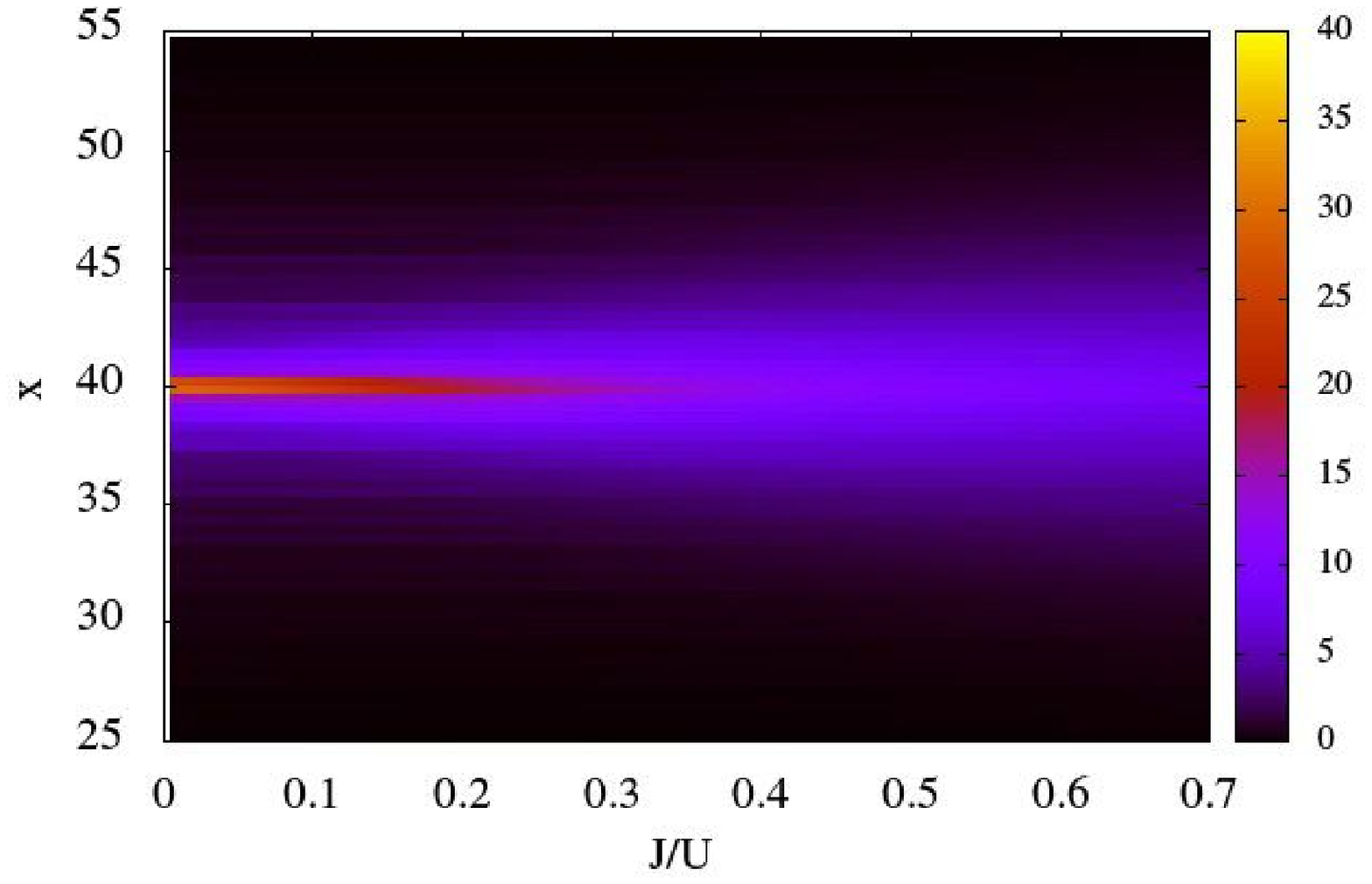}
 \caption{$\tilde{Z}_x$, the Fourier transform of the function $Z_q$,
   as a function of $x$ (the variable conjugate to $q$) and $J/U$ for
   a system of filling one.  Upper panel: exact diagonalization ($L=14$),
   lower panel: BWF-MC ($L=80$).  }
 \label{fig:heatmap}
\end{figure}

\section{Numerical Results}

\label{sec:results}

Fig. \ref{fig:ns_ED} shows the superfluid
weight for systems of different system sizes at filling one-half and
one.  The half-filled system is always superfluid.  The BWF-MC and ED
results are reasonably close in the case of half filling.  For integer
filling only the ED results are shown, since the BWF-MC are zero, as
shown above.  The ED results show no superfluid response for small
hopping values up to $J/U \sim 0.2$.  At that point the superfluid
weight starts to grow for the smallest system size. Larger system
sizes persist in an insulating state until larger values of $J/U$, but
all of them are noticeable by $J/U \approx 0.3$.  In the half-filled
system the energy as a function of $\Phi$ was found to be
discontinuous at a finite value of $\Phi$, indicating a level
crossing. No level crossing was found for filling one.  We note that
at filling one the gap closure (corresponding to the
Kosterlitz-Thouless transition)
occurs~\cite{Kuhner00,Ejima11,Zakrzewski08,Kashurnikov96a} at
$(J/U)_{KT}\approx 0.6$.  In particular, the ED calculation extended
by a renormalization group analysis by Kashurnikov and
Svistunov~\cite{Kashurnikov96a} gives $J_{KT}=0.608(4)$, nearly twice
the value where the superfluid weight begins to grow in
Fig. \ref{fig:ns_ED}.  Actually, for hopping value $J/U \sim
    0.3-0.5$, ED indicates a non-zero superfluid weight, but this is
    not a reliable result, once we are approaching a quantum
    transition, $(J/U)_{KT} \approx 0.6$, and the smallness of the
    system sizes, up to $L=14$, becomes highly manifest.

\begin{figure}[ht]
 \centering
 \includegraphics[width=6cm,keepaspectratio=true]{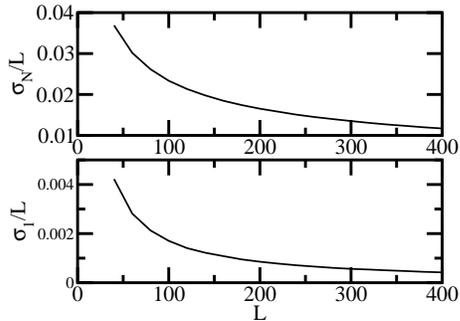}
 \caption{Standard deviation of the position (divided by the system size)
   vs. system size, for a system with $J/U=0.25$ at filling one.  Upper panel:
   total position for a bosonic system, lower panel: single-particle position
   with bosonic exchange moves turned off.}
 \label{fig:x}
\end{figure}
In Fig. \ref{fig:sigX2} the variance of the total position
$\sigma^2_N$ and its size scaling exponent $\gamma$ are shown.  The
size scaling exponent $\gamma$ is shown in the insets.  The size
scaling exponent $\gamma$ is known~\cite{Hetenyi19} to take the value
two in a closed gap (conducting) system, and the value one for an
insulating system.  We notice that in the upper panel of
Fig. \ref{fig:sigX2}, the exact diagonalization calculations bear out
these expectations, notwithstanding the limitations of small system
sizes.  The scaling exponent $\gamma$ is near the value one when $J/U$
is small, and increases to two around $J/U \sim 0.5$, in reasonable
agreement with the results of Kashurnikov and
Svistunov~\cite{Kashurnikov96a}.  For the BWF-MC calculations (lower
panel of Fig. \ref{fig:sigX2}), the size scaling exponent is near one,
in agreement with what is expected for an insulating phase.  As argued
in section \ref{sec:SFW} the superfluid weight is zero for the
Baeriswyl wave function at integer filling.  This conclusion is
corroborated by our variational results shown in the inset of the
lower panel of Fig. \ref{fig:sigX2}.

The upper panel of Fig. \ref{fig:heatmap} shows $\tilde{Z}_x$, the
discrete Fourier transform of $Z_q$ in the form of a color map $(J/U$
is the variable on the $x$-axis, the variable conjugate to
$\tilde{Z}_x$ is on the $y$-axis). The system is the unit filled one.
The interesting result on this figure is the peak, which occurs at
$x=7$, half the unit cell, which is where it should be for a
half-filled system with no spontaneous polarization.  The peak is
relatively constant until $J/U \approx 0.3$, then starts to decrease.
Again, this occurs at $J/U$ where the superfluid weight increases in
Fig.  \ref{fig:ns_ED}.  The lower panel of Fig. \ref{fig:heatmap} show
$\tilde{Z}_x$ as a function of $x$ and $J/U$ based on our variational
Monte Carlo results.  The similarity between the upper and lower
panels are striking: a sharp peak half-way through the supercell
exists near $J = 0$, which decreases around $J/U \approx 0.3$.
\textcolor{black}{An important difference between the two results is the
  scaling of the spread of the distributions, shown in the insets of
  Fig. \ref{fig:sigX2}.  Inspite of the qualitative similarity, the
  two distributions are indicative of different physical behavior.}

\begin{figure}[ht]
 \centering
 \includegraphics[width=6cm,keepaspectratio=true]{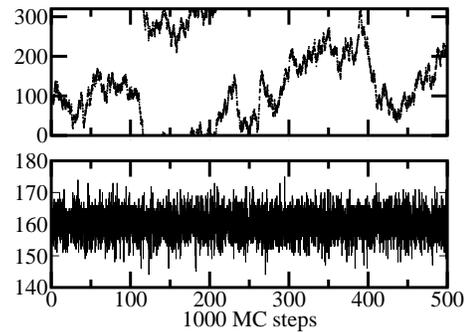}
 \caption{Upper panel: position of one particle on the lattice, lower panel:
   total position modulo the size of the system, for a system with $J/U=0.25$,
   $N=320$ particles, and size $L=320$ lattice sites.}
 \label{fig:xx1}
\end{figure}

In order to analyze the behavior of the superfluid weight (next
section), we present two more sets of results in Figs. \ref{fig:x} and
\ref{fig:xx1}.  The former shows the spread of the position as a
function of system size $J/U=0.25$ for two cases, both at filling one.
In the upper panel the total position is shown, indicating convergence
with system size, or insulating behavior.  In the lower panel, we show
the variance in the position of one particle, when bosonic exchange
moves are turned off (in other words, in a system of distinguishable
particles).  Fig. \ref{fig:xx1} shows two time series in the course of
the MC simulation, the upper panel for the position of a single
particle (with bosonic exchange moves), the lower panel for the total
position.  We see that even though the single particle delocalizes
over the whole lattice, the total position fluctuates around the
center of the unit cell.  We see that bosonic exchange delocalizes
single particles, but leaves the center of mass localized, fluctuating
around the center of the supercell.  Similarly, a superfluid weight
calculation which neglects the interference of particles would lead to
a finite superfluid response (we checked this).

\section{Discussion}

Anderson has argued~\cite{Anderson12}, based on a strong coupling
expansion, that the superfluid response of the BHM is finite at
integer filling.  Here, we show that his formalism is similar to our
strong-coupling expansion, and that his way of implementing the
Peierls flux does not allow for interference between particles.  The
finite superfluid weight is an artifact of this implementation.

We follow the steps of ~\cite{Anderson12}, but we a a system threaded
by a flux $\Phi$ at the outset.  For filling one in the strong
coupling limit $U\gg J$ and up to the leading order in $J/U$, the
many-body ground state is written as the product of non-orthogonal
bosons, the so called "eigenbosons",
\begin{equation}
\label{eqn:gs-anderson-phi}
|\Psi_A(\Phi)\rangle = \prod_{i=1}^L{{\hat b}^\dagger_i(\Phi)} |0\rangle
\end{equation}
where the "eigenbosons" are given by,
\begin{align}
\label{eqn:bc-transf-phi-L}
{\hat b}^\dagger_i(\Phi) & =  \frac{1}{M} \left[{\hat c}^\dagger_i + e^{-i\Phi} \frac{J}{2U} 
 {\hat c}^\dagger_{i+1} +  e^{i\Phi} \frac{J}{2U}{\hat c}^\dagger_{i-1} \right]
\end{align}
with $M=\sqrt{1+2(J/2U)^2}$.  Let us write down the first few terms of
Eq. (\ref{eqn:gs-anderson-phi}).  The zeroth order term is
\begin{equation}
\prod_i \hat{c}_i^\dagger | 0 \rangle
\end{equation}
which is equal to $|\infty\rangle$, the term zeroth order in $J/U$ in
Eq. (\ref{eqn:PsiSC1}).  The term first order in $J/U$ is
\begin{equation}
\frac{J}{2U}\sum_i [\exp(i\Phi) \hat{c}^\dagger_{i+1} + \exp(-i\Phi) \hat{c}^\dagger_{i-1}]
  \prod_{j\neq i} c_j^\dagger | 0 \rangle,
\end{equation}
which is exactly the first order term in Eq. (\ref{eqn:PsiSC1}).  For
this system we calculated the ground state energy, and showed that the
superfluid weight is zero.

The particular steps of Anderson, leading to the supefluid weight, are
as follows.  The bosonic states are Fourier transformed as,
\begin{eqnarray}
  \label{eqn:nonorthbosFT}
  \hat{b}_0 &=& \hat{c}_0 + \left(\frac{J e^{-i\Phi}}{2U}\hat{c}_1 +
  \frac{J e^{i\Phi}}{2U}\hat{c}_{-1} \right) \\ &=&
  \frac{1}{\sqrt{L}}\sum_k  \left[1 + (J/U) \cos(k +
    \Phi) \right] \hat{c}_k.  \nonumber
\end{eqnarray}
The kinetic energy is written
\begin{equation}
  \label{eqn:K}
  K = -\frac{J}{N}\sum_k \cos (k+\Phi)[1 + 2 (J/U) \cos \Phi \cos k + (J/U)^2 \cos^2 k]
\end{equation}
The second derivative of $K$ according to $\Phi$ will give a finite
contribution.  The issue is that in the Fourier transform of
Eq. (\ref{eqn:nonorthbosFT}), the Peierls phase is separated from the
rest of the system (in the words of Anderson a pair of lattice sites
is ``embedded'' in the system), and not allowed to interfere with the
Peierls phases of other hoppings.  When, in our BWF-MC calculations, a
Peierls phase is implemented in each of the propagators {\it
  independently,} without interference, a finite superfluid weight
also results.  Moreover, at small $J/U$ the two results give the same
scaling with $J/U$.

\section{Conclusion}  

\label{sec:conclusion}

In the context of solid $^4$He supersolidity was suggested by Kim and
Chan~\cite{Kim06} based on torsional oscillator experiments. It is
important to note that while the bosonic Hubbard model may give useful
hints into the behavior of solid helium, some aspects, for example
vacancies and dislocations, are not accounted for~\cite{Chan13}.  The
experiments of Kim and Chan were also brought in to question by the
suggestion~\cite{Beamish10} that their results could be due to quantum
plasticity.  Quantum plasticity is a phenomenon which can only be
defined in more than one dimension.

Our studies show that the Baeriswyl wave function can not produce a
finite superfluid weight.  Interestingly, the Fourier transform of the
polarization amplitude behaves very similarly to its counterpart
calculated by exact diagonalization.  Closer inspection however, for
example, the calculation of the size scaling exponent, still indicates
an insulating system. Therefore, the BWF is reliable up to $J/U
\approx 0.5$ not only for the ground state energy, but also for the
polarization amplitude in the insulating phase.

We also stressed the proper way to calculate the superfluid weight in
a variational context and commented on a calculation of Anderson based
on which a finite superfluid response was found for the integer filled
bosonic Hubbard model.  Our central result is that the ``usual''
formula of the superfluid weight, the second derivative with respect
to a momentum shifting Peierls phase, is not directly applicable in a
variational context.

\section*{Acknowledgments} 

We acknowledge financial support from TUBITAK under grant no.s 113F334
and 112T176.  BH was also supported by the National Research,
Development and Innovation Fund of Hungary within the Quantum
Technology National Excellence Program (Project Nr.
2017-1.2.1-NKP-2017-00001) and by the BME-Nanotechnology FIKP grant
(BME FIKP-NAT). BT acknowledges support from TUBA.  LMM acknowledges
J. Lopes dos Santos and J. V. Lopes for illuminating and stimulating
discussions.

\section*{Appendix:Strong coupling expansion based on the Baeriswyl wave function}

\label{sec:strong}

The BHM has been studied via a number of strong coupling
expansions~\cite{Freericks96,Elstner99,Damski06,Krutitsky08,Heil12}.
Such expansions in the case of the BHM are usually more difficult than
in the case of the fermionic Hubbard model, where the Heisenberg model
is obtained as the limiting case.

In the atomic limit, the 1D BHM ground state at filling one is given
by a full localized boson state
\begin{equation}
|\infty\rangle =\prod_{i=1}^L {\hat c}_i^\dagger|0\rangle.
\end{equation} 
Threading the system with a flux $\Phi$, using the fact that the state
$|\Psi_\infty\rangle$ is not $\Phi$ dependent, the BWF reads
\begin{equation}
|\Psi_B (\Phi) \rangle = \exp(-\alpha\hat{T}(\Phi))|\infty\rangle
\end{equation} 
where $\hat{T}(\Phi)$ is the first term in the r.h.s. of
Eq. (\ref{eqn:hml_Phi}).  Our first aim is to evaluate the variational
energy,
\begin{equation}  
\label{eqn:EBalphaPhi}
E_B(\alpha,\Phi) =  
\frac{\langle \Psi_B(\Phi)| {\hat H}(\Phi)  | \Psi_B(\Phi) \rangle}
{ \langle \Psi_B(\Phi)|\Psi_B(\Phi)\rangle},
\end{equation}
by performing a strong-coupling expansion ($J/U \ll 1$) via expanding
the kinetic projector and keeping the leading order terms in $J/U$ in all relevant quantities.
An important point is that successive applications of the kinetic
operator on $| \infty \rangle$ (a state with all sites occupied by one
particle) in expectation values of the type in
Eq. (\ref{eqn:EBalphaPhi}) should be such that one returns to the
state $\langle \infty|$.

Up to first order in $J/U$ the BWF is given by
\begin{align}
  \label{eqn:PsiSC}
|\Psi_B (\Phi) \rangle & = |\infty\rangle + \alpha J
|\Psi_d(\Phi)\rangle
\end{align}
where $\vert \Psi_d (\Phi)\rangle = \sum_{i=1}^L \left( e^{i\Phi}
\hat{c}^\dagger_{i+1} \hat{c}_i + e^{-i\Phi} \hat{c}^\dagger_{i}
\hat{c}_{i+1}\right) |\Psi_\infty\rangle$, a state which is a
superposition of states with occupation number $n_i$ being $0$ and $2$
for a pair of nearest neighboring sites and $1$ for all other sites.

The pair occupation, needed for the potential energy, within our
approximation is given by
\begin{equation}
\label{nd}
n_p = \langle \Psi_B(\Phi)| {\hat n}_i ( {\hat n}_i-1) |\Psi_B(\Phi)\rangle
     = 8 \alpha^2 J^2
\end{equation}
and the average of the potential energy by
\begin{equation}  
\label{EpalphaPhiL}
\langle \Psi_B (\Phi)|{\hat U}|\Psi_B (\Phi)\rangle = 8 \alpha^2 J^2 U L
\end{equation}  
where ${\hat U}$ is the second term in the r.h.s. of
Eq. (\ref{eqn:hml}).  The boson momentum distribution reads
\begin{equation}
n(k,\Phi) =  \langle \Psi_B(\Phi)| {\hat n}_k|\Psi_B(\Phi)\rangle =  1 + 8 \alpha J \cos(k+\Phi)\\ 
\label{nkalphaPhiL}
\end{equation}
where ${\hat n}_k = {\hat c}_k^\dagger {\hat c}_k$
and $\hat{c}^\dagger_{k} = \frac{1}{\sqrt{L}} \sum_{i=1}^{L} e^{-ik x_i}\hat{c}^\dagger_{i}$.
The above result is obtained by performing the calculations in momentum space
and exponentials $e^{\pm i(k+\Phi)}$ lead to the momentum
distribution to depend on $k+\Phi$. The average of the kinetic energy
is
\begin{equation}
\langle \Psi_B (\Phi)|{\hat T} (\Phi)|\Psi_B (\Phi)\rangle 
  =   \sum_{k} \epsilon_{k+\Phi}n(k,\Phi) = -8 \alpha J^2 L
\label{EkalphaPhiL}
\end{equation}
Note that both $\epsilon_{k+\Phi}$ and $n(k,\Phi)$ depend on $k+\Phi$,
therefore the kinetic energy is {\it{not flux dependent}} as expected,
since, at filling one, the product $\epsilon_{k+\Phi}n(k,\Phi)$ just
represents an overall shift in the Brillouin zone relatively to system
without flux. Optimizing the total energy at any flux leads to
$\alpha^* = \frac{1}{2U}$.  The optimal variational wavefunction is
given by
\begin{equation}
\label{eqn:PsiSC1}
|\Psi_B (\Phi,\alpha^*)\rangle  = |\infty\rangle + \frac{J}{2U} |\Psi_d(\Phi)\rangle
\end{equation}
and the ground state energy per site by
\begin{equation}
  E_B(\alpha^*) = - \frac{2 J^2}{U},
\end{equation}
where we dropped the $\Phi$ dependence from the notation. 
The inset in Fig. \ref{fig:ns_ED} compares this approximation to the results of
exact diagonalization for fourteen sites, and BWF-MC results for a
system of $L=160$.

As for the superfluid weight (Eq. (\ref{eqn:n_S})), we first express
the total current operator under a flux $\Phi$ within our
approximation as
\begin{eqnarray}
{\hat J} (\Phi) & = & \frac{ \partial {\hat T}(\Phi)}{\partial \Phi} =
2J \sum_{k} \sin(k+\Phi)\hat{c}^\dagger_{k} \hat{c}_{k},
\end{eqnarray}
and its average over $|\Psi_B (\Phi)\rangle$ is zero
\begin{eqnarray}
J_{\Phi}(\Phi) = 2J \sum_{k} \sin(k+\Phi) n(k,\Phi) = 0
\label{JkalphaPhiL}
\end{eqnarray}
indicating the insulating character of $|\Psi_B (\Phi)\rangle$ at
filling one.  The average of the current operator expressed by
Eq. (\ref{eqn:J_Phi}) reads
\begin{eqnarray}
  J_\Phi & = & 2J \sum_k \sin(k)n(k,\Phi) = -\frac{4J^2}{U}
  \sin\Phi L 
\end{eqnarray}
while the average of the kinetic energy, expressed by Eq. (\ref{eqn:T_Phi}) reads
\begin{eqnarray}
T_\Phi =  \sum_k \epsilon(k) n(k,\Phi) = -\frac{4J^2}{U} \cos\Phi L
\end{eqnarray}
Using Eq. (\ref{eqn:n_S}) $n_S=0$ for the BHM at
filling one in the strong coupling expansion, in agreement with the
previous ED and BWF-MC results.

For the 1D BHM at filling one up to order in $(J/U)^2$, the kinetic
and potential energy per site averages (at optimal variational
parameter), read $\langle{\hat T}\rangle = -\frac{4J^2}{U}$ and
$\langle {\hat U}\rangle = \frac{2J^2}{U}$, respectively. Physically,
this means that, through the kinetic exchange mechanism, the kinetic
energy gain is bigger than the on-site potential energy cost, in
perfect analogy with the Fermionic Hubbard Model (FHM) in the strong
coupling limit \cite{Fazekas99}.  Along the same lines, the pair
occupancy, Eq. (\ref{nd}), and momentum distribution for $\Phi=0$,
Eq. (\ref{nkalphaPhiL}), read  as
\begin{eqnarray}
n_p & = & \frac{2J^2}{U^2}
\end{eqnarray}
and
\begin{equation}
\label{eq:nkSCBWF2}
n(k) = 1 - \frac{2 \epsilon_k}{U}
\end{equation}
also in perfect analogy with the FHM in the strong coupling limit,
except, obviously, for the absence of the spin-spin term
\cite{Fazekas99}. It is also worth mentioning that the momentum distribuition for $\Phi=0$,
Eq. (\ref{eq:nkSCBWF2}), is in agreement with that of Ref. \cite{trivedi2009}.

\end{document}